\documentclass[12pt]{article}

\usepackage{xr}
\usepackage{hyperref}
\usepackage{scicite}
\usepackage{graphicx}
\usepackage{caption}
\usepackage{times}
\usepackage{xcolor}
\usepackage{gensymb}
\usepackage{amsmath,amssymb}
\usepackage{graphicx}
\usepackage{hyperref}
\newenvironment{sciabstract}{%
\begin{quote} \bf}
{\end{quote}}

\newcounter{lastnote}

\title{Raman scattering of phonon polaritons under nanoscale confinement: the role of structure and environment}

\author
{George Zograf,$^{1,\dagger,\ddagger}$ Bet$\textrm{\"u}$l K$\textrm{\"u}\textrm{\c{c}}\textrm{\"u}$k$\textrm{\"o}$z,$^{1,\ddagger}$ \\ 
Oleg Kotov,$^{1,\P}$ Naveen Shetty,$^{2}$ \\ Lunjie Zeng,$^{1}$ Andrew B. Yankovich,$^{1}$ Alok Ranjan,$^{1}$ \\
Avinas N. Shaji,$^{4}$ Erik Lind,$^{4}$ Tomasz J. Antosiewicz,$^{1,3}$ \\
Eva Olsson,$^{1}$ Samuel Lara-Avila,$^{2}$ Timur O. Shegai$^{1,^\ast}$\\
\\
\normalsize{$^{1}$Department of Physics, Chalmers University of Technology, Gothenburg, 412 96, Sweden}\\
\normalsize{$^{2}$Department of Microtechnology and Nanoscience, Chalmers University of Technology,}\\
\normalsize{Gothenburg, 412 96, Sweden}\\
\normalsize{$^{3}$Faculty of Physics, University of Warsaw, Pasteura 5, 02-093 Warsaw, Poland}\\
\normalsize{$^{4}$Department of Electrical and Information Technology, Lund University, 221 00 Lund, Sweden}\\
\normalsize{$^\ddagger$These authors contributed equally to this work}\\
\normalsize{$^\ast$To whom correspondence should be addressed; E-mail:  timurs@chalmers.se}\\
\normalsize{$^\dagger$Current address: School of Electrical and Electronic Engineering, Nanyang}\\
\normalsize{Technological University, 639798, Singapore} \\
\normalsize{$^\P$Current address: Departamento de Física Teórica de la Materia Condensada and}\\
\normalsize{Condensed Matter Physics Center (IFIMAC), Universidad Autónoma de Madrid,} \\
\normalsize{Madrid, 28049, Spain}
}

\date{}

\begin{document} 

% Double-space the manuscript.

\baselineskip24pt

% Make the title.

\maketitle 

% Place your abstract within the special {sciabstract} environment.

%Phonon polaritons are promising for nanoscale light manipulation, energy confinement, and light up-conversion because of strong field localization and long lifetimes in the mid-infrared spectral range. 

\newpage

%(composed of silicon carbide or gallium nitride)

\begin{sciabstract}
Strong light-matter coupling gives rise to polaritons -- quasiparticles that combine both photonic and material characteristics. Here, we show that polar nanocrystals exhibit structure- and environment-dependent Raman scattering, enabled by their hybrid phonon polariton nature. Such dispersive behavior enables refractive index sensing in the mid-infrared range via visible-wavelength inelastic spectroscopy and draws parallels with molecular systems under vibrational strong coupling. Crucially, Raman scattering appears only under nanoscale confinement of phonon polaritons. For optimal structures, this leads to self-hybridization between localized phonon modes and surface phonon polaritons hosted by the same nanoparticle.  

%Transverse modes are observed in both Raman and infrared spectroscopy, whereas longitudinal ones appear exclusively in the infrared, indicating that Raman excitation requires phonon-polariton localization on a scale smaller than the excitation wavelength. 

%%%% removed from abstract, but might return
% inherited from their photonic component

% this is important, maybe part of the introduction in conference talks or papers:
%The role of matter consists in enabling inelastic light scattering (Raman), while the role of light lies in its wavelength-dependent dispersion, which defines sensitivity to structural and environmental variations. In other words, matter provides the Raman activity, while light provides the structural dispersion.

\end{sciabstract}

\paragraph*{Raman scattering of polaritons and vibrational strong coupling.}

Raman scattering originates from inelastic scattering of photons by material excitations, such as vibrations in molecular compounds or optical phonons in solids~\cite{HayesLoudon1978raman}. In molecules, it is generally not sensitive to variations in macroscopic parameters such as the electromagnetic environment or the angle of incidence. In polar solids, however, Raman scattering can involve not only pure vibrations but also hybrid light-matter quasiparticles -- phonon polaritons~\cite{huang1951lattice}. In this case, the Raman process may become dependent on the angle of incidence relative to both the crystal lattice and the wave-vector of the scattered light. This was first demonstrated by Henry and Hopfield~\cite{henry1965raman}, who observed angle-dependent Raman scattering by phonon polaritons in bulk gallium phosphide (GaP) crystals at frequencies below the transverse optical (TO) phonon. These experiments required a near-forward scattering geometry to satisfy momentum conservation. Later, similar findings have been reported in other polar crystals, such as GaN~\cite{davydov1997raman,torii2000raman,irmer2013phonon} and GaSe~\cite{bergeron2023probing}. In the latter case, Raman scattering exhibits thickness-dependent behavior arising from dispersion engineering of surface phonon polaritons (SPhP).

% nanostructuring and surface phonon polaritons
% (similarly to the localized surface plasmon polaritons in noble metal nanoparticles)

\begin{figure}%[ht!]
    \centering
    \includegraphics[width=0.9\linewidth]{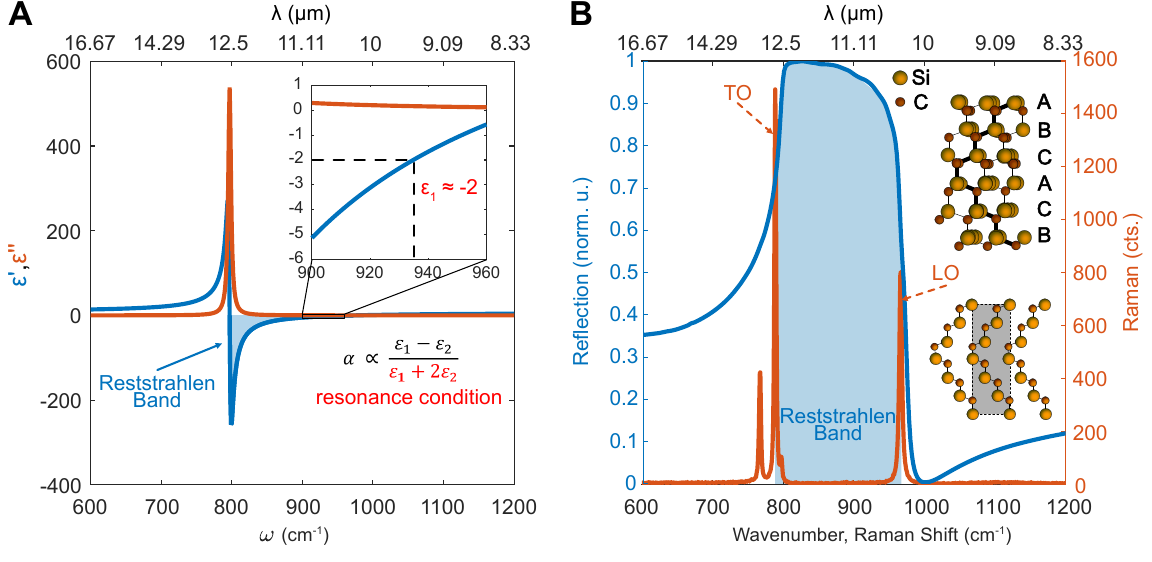}
    \captionsetup{font=scriptsize}
    \caption{\textbf{Optical constants, Raman and infrared spectra of bulk SiC.} \textbf{A}, Analytical permittivity of SiC: real (blue) and imaginary (red) part, reproduced from Palik handbook~\cite{choyke1997silicon}. Inset demonstrates the zoom in part where the resonance condition is reached for a subwavelength sphere, $\varepsilon_1\simeq -2$ is valid for air medium. \textbf{B}, Infrared reflection (blue line, left axis) and Raman scattering (red line, right axis) of 6H-SiC crystal. Insets show the schematic of the crystal lattice.}
    \label{fig:SiC}
\end{figure}

Nanostructuring lifts the near-forward scattering constraint, as first shown by Hayashi \textit{et al.} in GaP nanocrystals~\cite{hayashi1982raman}. These observations ultimately led to the formulation of three conditions for Raman scattering by SPhPs~\cite{hayashi1984optical,hayashi1985raman}: (\textbf{1}) the SPhP is observed within the Reststrahlen band, where the real part of the material's dielectric function is negative, $\varepsilon' < 0$ (Figure~\ref{fig:SiC}), (\textbf{2}) the relative intensity of the SPhP peak increases for smaller particles compared to the intensity of the TO phonon peak, and (\textbf{3}) the frequency and intensity of the SPhP peak should be sensitive to the dielectric constant of the surrounding medium. Following these criteria, nanostructured GaP, GaN, GaAs, InAs, and SiC have been reported to exhibit Raman scattering of SPhPs~\cite{hayashi1982raman,sekine2017surface,rybchenko2021,moller2011polarized,digregorio1994raman}. In contrast, contemporary phonon-polariton materials such as hexagonal boron nitride (hBN) and molybdenum trioxide ($\alpha$-MoO$_3$)~\cite{li2018infrared,hu2020topological} lack Raman-active bulk phonon polaritons due to crystal inversion symmetry, which enforces mutual exclusion between Raman and infrared modes. Thus, Raman scattering of bulk phonon polaritons requires a combination of a Reststrahlen band, Raman-active longitudinal optical (LO) and TO phonons, and second-order nonlinearity (lack of inversion symmetry), criteria met, \textit{e.g.}, by GaP, GaN, GaSe, GaAs, InAs, SiC, AlN, ZnO, HfS$_2$~\cite{kowalski2025ultraconfined}, r-BN, and other polar crystals. In nanostructures, the particle size, $D$, must additionally be smaller than the excitation wavelength, $\lambda_{\mathrm{exc}}$~\cite{hayashi1982raman,hayashi1984optical,hayashi1985raman,digregorio1994raman}.

In an alternative perspective, phonon polaritons may be interpreted as light-matter hybrids in the vibrational strong coupling (VSC) regime~\cite{shalabney2015coherent,barra2021microcavity,hertzog2021enhancing}. Similarly to SPhPs in solids, VSC in cavity-molecule systems is expected to generate a dispersive Raman signal that differs from that of uncoupled molecules. Shalabney \textit{et al.}~\cite{shalabney2015enhanced} demonstrated precisely that in a Fabry-Pérot microcavity loaded with polyvinyl acetate molecules supporting infrared- and Raman-active vibrations under VSC. Subsequently, Raman scattering from plasmonic terahertz cavities coated with a layer of polar CdS nanocrystals in the strong coupling regime revealed Raman bands at the upper and lower polariton branches~\cite{jin2018reshaping}. A modification of Raman scattering of a Fabry-Pérot microcavity in the VSC regime with C=O vibrations in polymethyl methacrylate was also reported~\cite{ahn2020raman}. Furthermore, Raman scattering was used as a local probe of ultrastrong coupling in VSC systems~\cite{arul2023raman}. However, several recent works have reported no significant differences between the Raman signal of VSC systems and that of bare molecules~\cite{takele2021scouting,verdelli2022chasing,menghrajani2022probing,cohn2023spontaneous}. In addition, several theoretical works demonstrated VSC signatures in Raman spectra but did not confirm the enhancement reported in the original experiments~\cite{del2015signatures,strashko2016raman}. 

% similarity between VSC and SPhPs. This similarity is intriguing, but we do not use it in the text, and it sounds somewhat speculative. Perhaps one should expand this in the experiment and even include this in the title to better reflect the point.
%Insights from the phonon polariton literature could improve our understanding of Raman scattering of VSC systems. 

It is of interest to bridge the observations of Raman scattering by phonon polaritons to the recent developments in VSC. The SPhP–VSC comparison becomes even more compelling by recognizing that phonon polaritons can be viewed as bulk Hopfield polaritons~\cite{hopfield1958theory} and that in nanostructured systems, they transform into self-hybridized polaritons, recently reported in a wide range of material platforms~\cite{munkhbat2018self,canales2021abundance,canales2023perfect,dirnberger2023magneto,canales2024self,ziegler2025electrical}. Furthermore, the optical second-order nonlinearity in Raman-active SPhP systems links them to light up- and down-conversion and molecular optomechanics~\cite{chen2021continuous,xomalis2021detecting,niemann2024spectroscopic}.

%, making the direct comparison of Raman scattering of SPhPs with VSC systems especially relevant

%Moreover, optical second-order nonlinearity may link the Raman scattering of SPhPs to efficient up- and down-conversion processes, analogous to sum-frequency generation (up-conversion) and its inverse variant -- spontaneous parametric down-conversion directly involving phonon polaritons and their molecular optomechanical analogs~\cite{chen2021continuous,xomalis2021detecting,niemann2024spectroscopic}.

With this in mind, we note that SiC presents a remarkable phonon polariton platform~\cite{hillenbrand2002phonon}, featuring a wide Reststrahlen band with $\omega_{LO}-\omega_{TO}$ splitting of $\sim$ 176 cm$^{-1}$, which is equivalent to the bulk coupling strength of $g = \sqrt{\omega_{LO}^2 - \omega_{TO}^2}/2 = 278$ cm$^{-1}$ and normalized coupling strength $\eta=g/\omega_{TO}=0.35$ -- a hallmark of ultrastrong light-matter coupling~\cite{canales2021abundance,vicentini2025real}, Raman-active LO and TO phonons (Figure~\ref{fig:SiC}b), and second-order optical nonlinearity~\cite{lundquist1995second}. These properties satisfy the Raman scattering criteria for SPhPs, as evidenced in both SiC nanocrystals~\cite{sasaki1989optical,digregorio1994raman} and top-down-fabricated SiC resonators~\cite{caldwell2013low,beechem2019influence}. Furthermore, from a material perspective, SiC is chemically inert, mechanically robust, and possesses a high melting point ($T_{\rm m}>1000^{\circ}$). Motivated by these unique properties, we focus here on a range of SiC nanostructures, with the primary goal of elucidating the structure- and environment-dependent Raman response of SPhPs, correlating it with infrared spectroscopy, and strengthening the analogy to VSC. More generally, we find that the material component of the polaritonic system is responsible for enabling Raman scattering, while the photonic component of the polaritonic system defines its sensitivity to structural and environmental variations.

\paragraph*{Surface phonon polaritons in SiC nanopillars.}

The dielectric function of SiC is approximated by the following:
\begin{equation}
    \varepsilon_{SiC}(\omega) = \varepsilon_\infty\left( 1 + \frac{\omega_{LO}^2 - \omega_{TO}^2}{\omega_{TO}^2-\omega^2-i \omega \gamma}\right) 
    \label{Eq:eps_SiC}
\end{equation}
with the following parameters: $\varepsilon_\infty = 6.7$, $\omega_{TO}$ = 793~cm$^{-1}$, $\omega_{LO}$ =~969 cm$^{-1}$, and $\gamma$ = 4.76~cm$^{-1}$~\cite{choyke1997silicon} (material anisotropy is neglected for simplicity). The Reststrahlen band occurs at $\omega_{LO} > \omega > \omega_{TO}$, in the region where $\varepsilon'(\omega) < 0$ and thus SPhPs are supported, similar to conventional metal-based plasmonics (Figure~\ref{fig:SiC}A). Both TO and LO phonons are Raman-active in SiC (Figure~\ref{fig:SiC}B), which enables the observation of SPhPs in SiC through infrared spectroscopy and Raman scattering.

%  The negative dielectric function within the Reststrahlen band results in a significant reflection in the infrared range (Figure 1B).

\begin{figure}%[ht!]
    \centering
    \includegraphics[width=0.99\linewidth]{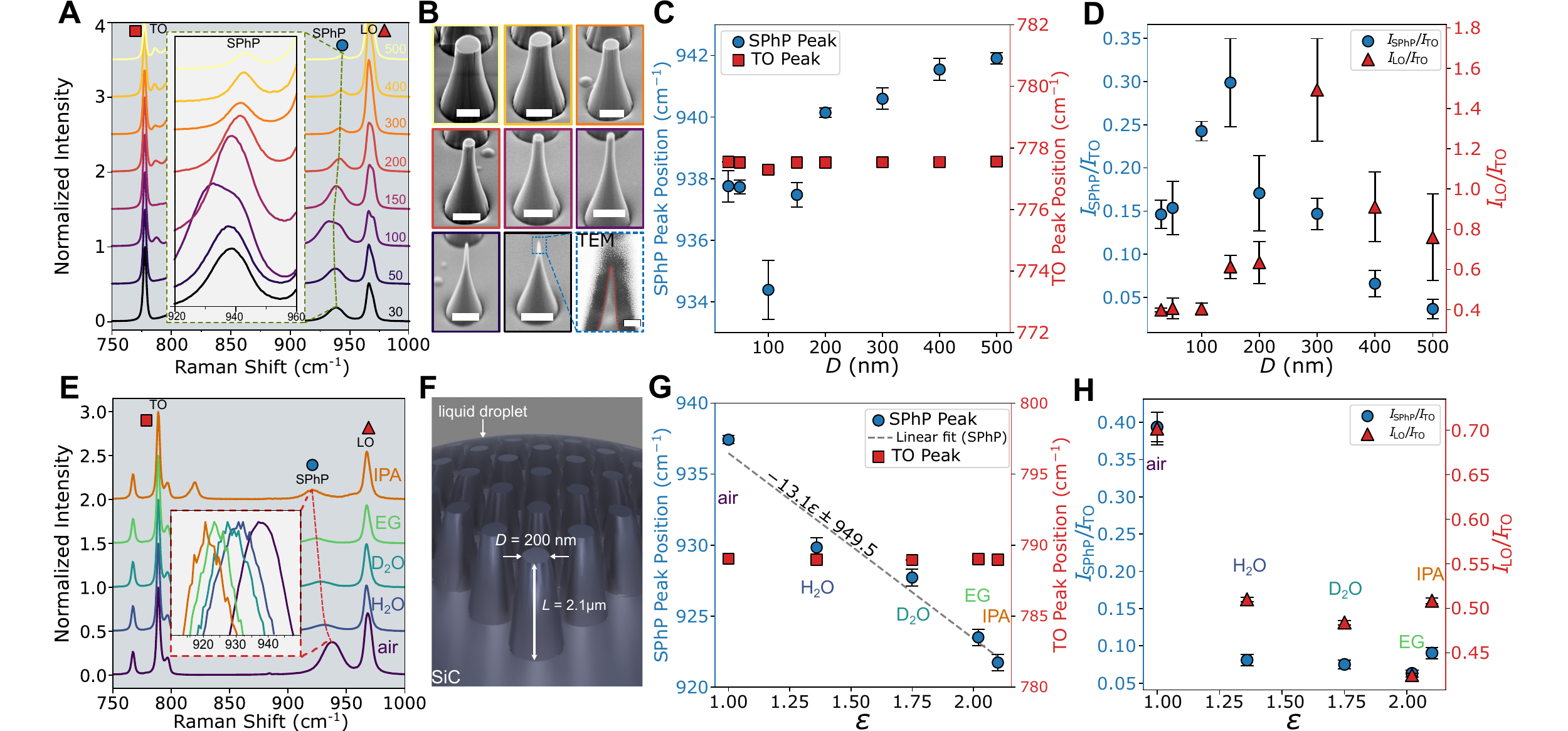}
    \captionsetup{font=scriptsize}
    \caption{\textbf{Raman scattering by surface phonon polaritons in SiC nanopillars: the effect of structure and environment.} \textbf{A}, Typical experimental Raman scattering spectra for different diameters of nanopillars. Each spectrum has a removed baseline signal and normalized over TO peak intensity. The offset of 0.5 between neighboring spectra along the $y$-axis is artificial for visual perception.Inset shows the enlarged view of SPhP resonance peak shift. \textbf{B}, Tilted SEM images of the SiC nanopillars. Scale bars are 500 nm. The color matches the corresponding plots in \textbf{A}. The blue dashed line indicates a close-up cross-sectional ADF STEM image of the $\sim$ 10~nm diameter pillar. Scale bar for the STEM image is 50~nm. \textbf{C}, Statistically averaged Raman scattering fitting of the TO and SPhP peak positions. Error bars are from the standard deviation of the mean value. \textbf{D}, The ratio between SPhP or LO intensity and a TO intensity as a function of the diameter. Error bars are statistical error bars from fitting of 9 separate disks. \textbf{E}, Raman scattering spectra of a dense array of SiC $D=200$~nm nanopillars for different surrounding media -- air, water H$_2$O, heavy water D$_2$O, ethylene glycol (EG), and isopropanol (IPA). Similar normalization and offset as in \textbf{A}. Inset: Enlarged view of the SPhP resonance peak shift observed for different surrounding media. \textbf{F}, Schematic of the experiment where the sample was covered homogeneously with a liquid droplet. \textbf{G}, Raman scattering fitting of the TO and SPhP peak positions as a function of the real part of the media permittivity $\varepsilon'$ at 930~cm$^{-1}$. The dashed line is the linear fit of the SPhP peak spectral position. Error bars represent the 0.95 confidence interval. The dielectric function of heavy water at 10.75 $\mu$m is adopted from Ref.~\cite{bertie1989infrared}. \textbf{H}, The ratio between SPhP or LO intensity and a TO intensity as a function of the media permittivity. Error bars represent the 0.95 confidence interval.}
    \label{fig:environment}
\end{figure}

%  \red{TS: As it turns out, the LO/TO ratio is relative weakly discussed in the text, therefore, I suggest considering removing it altogether from Figures 2-3 and from the text (it is too weak to mention). Let's discuss it together though.}

% , due to limitations in the etching process

We fabricated SiC pillars using both 6H- and 4H-SiC (pillar height, $L$ = 1 -- 2~$\mu$m) with variable top diameter, $D$, in the range of $\sim$ 10 -- 500~nm directly on a SiC substrate. The 6H- and 4H-polytypes of SiC yielded similar results, except for an additional TO mode in the 6H case (see Figure S1 for 4H-6H comparison). $L$ was kept constant throughout the sample. Figure~\ref{fig:environment}B shows the tilted-view scanning electron microscopy (SEM) images. Note that the shape of the pillar resembles a truncated cone rather than a perfectly vertical cylinder (see Methods). The Raman spectra of SPhPs reflect this complexity through \textit{e.g.} the broadened linewidth of the corresponding resonances in comparison to the TO phonon linewidth. Such peculiar shapes arise from the nanofabrication process, enabling the creation of ultrasharp needle-like pillars with $D < 10$ nm, as illustrated in the bottom-right SEM and annular dark field (ADF) scanning transmission electron microscopy (STEM) images in Figure~\ref{fig:environment}B. The corresponding Raman spectra show the appearance of SPhP peaks within the Reststrahlen band in the $\sim$ 935 -- 943~cm$^{-1}$ range, in addition to LO and TO phonons present in the bulk crystal (Figure~\ref{fig:environment}A). For a flat unpatterned SiC substrate, SPhPs are \textit{not} observed (Figure~\ref{fig:SiC}B).

% at $\sim$~969~cm$^{-1}$ and $\sim$~775~cm$^{-1}$, respectively

The SPhP peaks consistently shift to lower frequencies (\textit{i.e.} redshift) as $D$ decreases (Figure~\ref{fig:environment}C, with the exception of the two smallest pillars, which have sharp needle-like shapes), confirming its polaritonic nature. Indeed, the diameter-dependent dispersion indicates the hybrid light-matter character of the observed signal, with the photonic component responding to structural variations. This is analogous to VSC, as the diameter-dependent dispersion observed here (Figure~\ref{fig:environment}C), is reminiscent of the angle- or cavity-thickness-dependent dispersion characteristic of VSC systems~\cite{shalabney2015coherent}. In contrast, the TO peak is invariant with respect to nanopillar size, consistent with its purely vibrational (non-polaritonic) origin.

The Raman intensity of the SPhP mode relative to that of TO phonons decreases as the particle size increases. This SPhP/TO ratio is non-monotonic for $D$ = $\sim$ 10 -- 500~nm, peaking at $D=150$~nm (Figure~\ref{fig:environment}D), indicating that only relatively small nanocrystals exhibit Raman activity. The spectra were collected from individual SiC pillars within 2 $\mu$m pitch arrays, using a focused $\lambda_{\mathrm{exc}}$ = 532~nm laser beam, which allowed the excitation to be localized on a single pillar (Figure S2). Notably, the Raman signal drops for diameters $D < 150$~nm, likely due to the increasingly complex shape of the nanopillars and smaller overall nanoparticle volume, violating the second Raman SPhP condition (\textbf{2}), a point to which we return later. Additionally, for large $D=500$~nm particles, the SPhP/TO ratio is relatively low and is expected to decrease further with increasing diameter, suggesting that in the $D \rightarrow \infty$ limit, the signal vanishes, as observed in continuous SiC films (Figure~\ref{fig:SiC}B). Thus, our experiments demonstrate that SPhPs must be localized on a scale smaller than the excitation wavelength, $D < \lambda_{\mathrm{exc}}$, to be efficiently excited. By analogy with VSC experiments~\cite{takele2021scouting,verdelli2022chasing,menghrajani2022probing,cohn2023spontaneous}, we anticipate that in large translationally invariant systems (\textit{e.g.}, planar Fabry–Pérot microcavities), the Raman signal may vanish, unless structural roughness exists on a scale smaller than the excitation wavelength, or the signal is collected in the near-forward scattering geometry.

%(\red{GZ: Shall we emphasize here that it matches the Hayashi criterion number 2? TS: Maybe with different wording. It turns out that 'Hayashi criteria' is not a generally accepted term in spectroscopy, so we should probably avoid calling them that explicitly.})

The Raman intensity of the LO modes varies with particle size, peaking at $D = 300$ nm, relative to the TO phonons. This suggests a possible redistribution of oscillator strength between SPhP and LO modes. Additionally, optical resonances of the SiC structures in the visible range, overlapping with both the excitation and Raman-shifted wavelengths, may contribute to this behavior.

% and despite this contradicting the SPhP Raman condition, which requires a continued increase in signal with decreasing size

%In particular, the applicability of the $D < \lambda_{\mathrm{exc}}$ condition, known for SPhPs, to VSC systems is in question, potentially explaining recent observations~\cite{takele2021scouting,verdelli2022chasing,menghrajani2022probing}. 

%An additional analogy with VSC is the diameter-dependent dispersion observed here (Figure~\ref{fig:environment}B,C), which is reminiscent of the angle- or cavity thickness–dependent dispersion characteristic of VSC systems~\cite{shalabney2015coherent}.    

To further investigate the nature of SPhPs' Raman scattering, we performed anti-Stokes Raman measurements and compared the resulting Stokes/anti-Stokes intensity ratios to those predicted by the Boltzmann distribution. These measurements were performed on dense SiC nanopillar metasurfaces to collect a sufficient amount of the anti-Stokes signal. The results confirm that the observed distribution is thermal, consistent with expectations for incoherent phonon populations (Figure S5).

%\begin{figure}%[ht!]
%    \centering
%    \includegraphics[width=0.75\linewidth]{Figs/TOC_1.0.pdf}
%    \caption{\textbf{Concept of the size effect plasmon-polariton Raman scattering in SiC nanopillars} SiC nanopillars with different diameters which allows to tune the Raman mode related to plasmon-polariton mode in mid-IR range.}
%    \label{fig:TOC}
%\end{figure}

%The substrate is SiO$_2$/Si, with SiO$_2$ thickness of about 200 nm. This limits Raman scattering to back-scattering configuration. Pure glass is not suitable as a substrate, because it is nearly intransparent in the mid-infrared range and might affect phonon polaritons drastically. On the other hand, pure Si is probably also not so good as a substrate, because (a) it reduces the collection efficiency of Raman photons and (b) it has a high refractive index also in the mid-infrared range, again causing disturbance of the polaritonic eigenstates.

\begin{figure}%[ht!]
    \centering
    \includegraphics[width=0.99\linewidth]{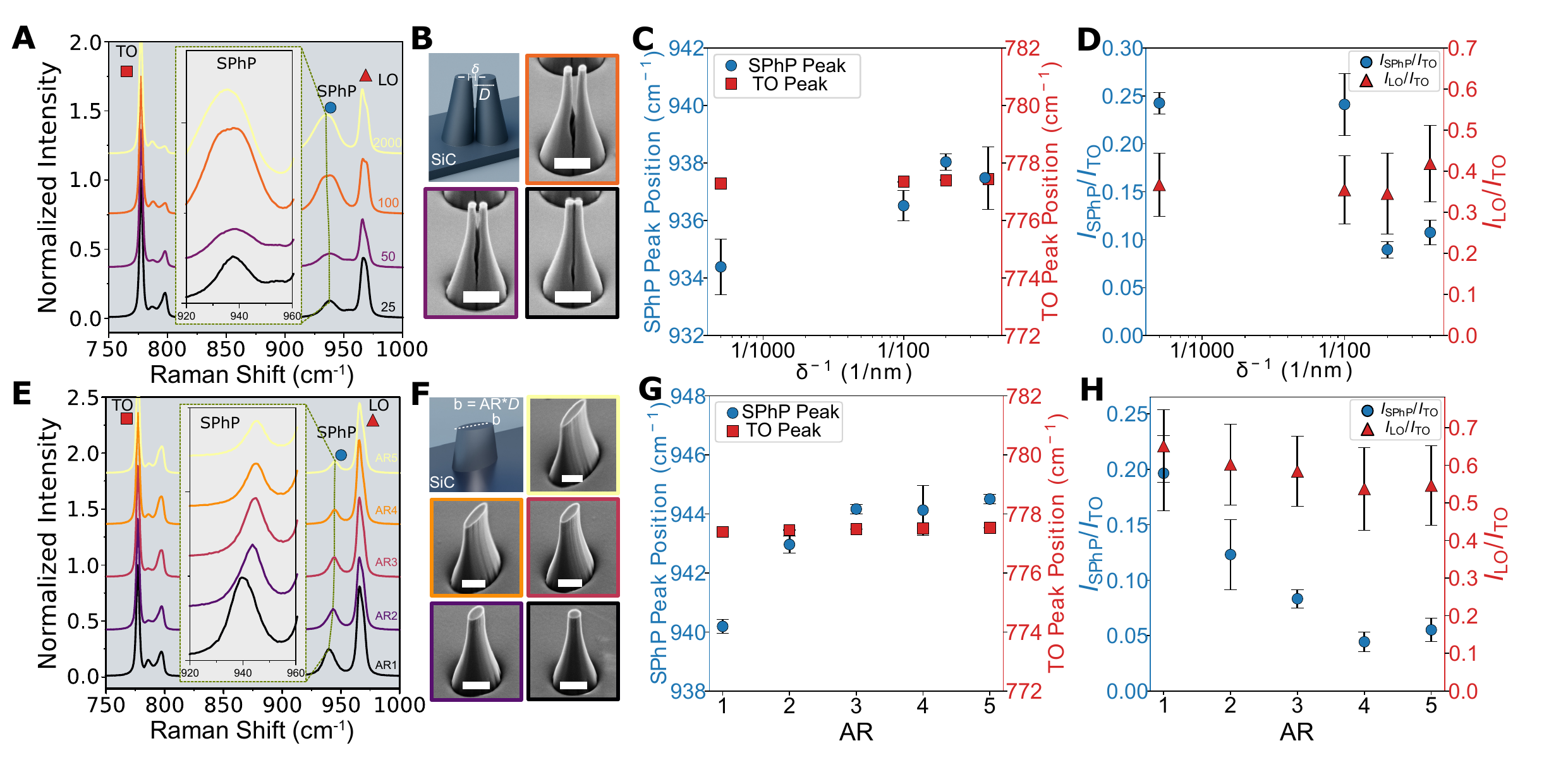}
    \captionsetup{font=scriptsize}
    \caption{\textbf{Raman scattering by surface phonon polaritons in SiC dimer and elliptical nanopillars.} \textbf{A}, Typical Raman spectra of the dimer nanopillars of $D$ = 100 nm diameter and 25, 50, 100, and 2000 (single pillar) nm gap size ($\delta$). Normalization and offset are made similar to Figure~\ref{fig:environment}\textbf{A}. Inset shows the enlarged view of SPhP resonance peak shift with different gap sizes. \textbf{B}, Schematic of the dimensions of the structure and typical SEM images of the dimers. Scale bars are 500 nm. Boxes colors correspond to frames color from \textbf{A} $\delta$ = 25, 50, 100 nm. \textbf{C}, Statistically averaged over 9 dimers per gap reference TO Raman peak (red squares) and SPhP peak (blue circles) positions as functions of gap size $\delta$. \textbf{D}, The relative intensity of SPhP (blue circle) and LO (red triangles) peaks to TO peak for each gap size, averaged over 9 dimers per gap. \textbf{E}, Typical Raman spectra of the elliptical nanopillars of $D$ = 200 nm diameter and 1 to 5 aspect ratio (AR) of the corresponding color of the frames in \textbf{F}. Normalization and offset are made similarly to Figure~\ref{fig:environment}\textbf{A}. Inset shows the enlarged view of SPhP resonance peak shift with different aspect ratios. \textbf{F}, A graphical image schematically showing the AR correspondence to the diameter and tilted SEM images of the SiC nanopillars. Scale bars are 500 nm. Colors of the frames correspond to spectra color AR = 1 -- 5. \textbf{G}, Statistically averaged over 9 elliptical pillars per AR value reference TO Raman peak position (red squares) and SPhP peak position (blue circles) as functions of AR. \textbf{H}, Relative intensity of SPhP and LO peaks to TO peak intensity for each AR value, averaged over 9 ellipsoids per AR.}
    \label{fig:shape}
\end{figure}

%  \red{A scientific question: why do anisotropic particles not produce "double-peaked" Raman spectra, with one peak along the long and the other along the short axis of the ellipsoid? I believe Raman measurements were performed without a polarizer. Maybe a high-resolution Raman mapping of longer AR ellipsoids should be done to verify the effect. BK: We have tried high-resolution mapping for elliptical particles, and there is no double peak that we can conclude.}

%In case this influence is observed, this will further confirm the SPhP nature of the signal.

\paragraph*{Changes in environment.} Following the third Raman SPhP condition (\textbf{3}), we now aim to verify the influence of the dielectric environment on the Raman spectra of SiC nanopillars, schematically illustrated in Figure~\ref{fig:environment}F. To increase the signal, for this experiment, we used dense SiC metasurfaces consisting of $D$ = 200 nm pillars with a pitch of 600~nm (see Methods). The spectra were measured in air and in four different liquids -- water (H$_2$O), heavy water (D$_2$O), ethylene glycol (EG), and isopropanol (IPA). The SPhP Raman peak shows a significant and nearly linear redshift for these liquids (up to $\sim$~15~cm$^{-1}$), further confirming its polaritonic nature. In contrast, the LO and TO phonon peaks remain largely unaffected by changes in the surrounding environment (Figure~\ref{fig:environment}E,G).

The SPhP/TO and LO/TO ratios, shown in Figure~\ref{fig:environment}H, experience a pronounced, nearly 4-fold, reduction for the former when the sample is immersed in liquids. This supports the localization requirement, $D < \lambda_{\rm exc} / n_{\rm vis}$, as the excitation wavelength is shortened in the medium by its refractive index in the visible range, $n_{\rm vis}$, while $D$ remains fixed (200 nm in this case).

The environmental sensitivity of Raman scattering by SPhPs calls for parallels with plasmonic refractive-index sensing in the visible -- near-infrared range~\cite{mayer2011localized}, with the key distinction of the former being an inelastic scattering. Unlike plasmonics, the Raman SPhP approach allows measuring the signal through environmental changes in the mid-infrared range -- specifically, in the molecular fingerprint region, by using visible or near-infrared optics. This is potentially advantageous over existing plasmonic sensing platforms, particularly in terms of high variations of refractive indices of organic compounds in the fingerprint region. Furthermore, specific molecular groups -- such as C-O-H -- can dominate the signal in the $\sim$ 930 -- 940 cm$^{-1}$ range, resulting in enhanced sensitivity to compounds containing these functional groups. An example of this can be seen in EG and IPA in Figures~\ref{fig:environment}E,G. Additionally, the mid-infrared spectral range supports a remarkably wide span of dielectric function values, from approximately –1000s for metals (\textit{e.g.} Au, Ag, Al, etc.) to around +100s for phonon polariton materials (\textit{e.g.} hBN, MoO$_3$, etc.). This range far exceeds the typical dielectric function values accessible in the visible to near-infrared spectrum, which are utilized in conventional plasmonics-based refractive index sensing. Raman SPhP approach thus combines the best of both worlds: the strong, molecule-specific refractive index contrast of the mid-infrared range with the practical advantages of visible-wavelength spectroscopy.

\paragraph*{Dimers and ellipsoids.} An alternative strategy for tailoring the electromagnetic environment involves structuring the material into more complex geometries. In this study, we focus on nanopillar dimers, anisotropic shapes such as elliptical nanopillars, and nanopillar heptamers -- structures that are well-studied in plasmonics. We anticipate that plasmonic-inspired designs can facilitate the formation of phonon-polariton ``hot spots`` and Fano resonances. A key objective is to investigate how these effects influence Raman scattering.

%The impact of nanopillar gap size and aspect ratio is particularly pronounced in experiments. 
%This trend contrasts with the red shift observed for decreasing particle size in individual pillars (Figure~\ref{fig:environment}). 

For nanopillar dimers, we investigate several gap sizes, $\delta$ = 25, 50, and 100 nm, while keeping $D$ and $L$ constant ($D=100$ nm and $L$ = 1 -- 2 $\mu$m); data are shown in Figure~\ref{fig:shape}A-D. As the gap between nanopillars decreases, we observe a blue shift of the SPhP resonance by approximately 3 -- 3.5~cm$^{-1}$ (\ref{fig:shape}C), accompanied by a reduction in the SPhP-to-TO intensity ratio (Figure~\ref{fig:shape}D). The response of isolated pillars complements this set at $\delta$ = 2000 nm, corresponding to the pitch used in Figure~\ref{fig:environment}. We note that the dimers possess a complex truncated cone shape, resulting in a shared base that forms a continuous structure, rather than a pair of perfectly vertical non-overlapping cylinders (Figure~\ref{fig:shape}B).

% Thus, here, a smaller gap may mean an overall bigger structure, whose SPhP peak should indeed be shifted to the blue (see Figure 1B).

To further illustrate the effect of nanopillar shape, we fabricated elliptical nanopillars with varying aspect ratios (AR), as shown in Figure~\ref{fig:shape}F. We observe that increasing the aspect ratio leads to a blue shift of the SPhP Raman peak, while the TO peaks remain nearly unchanged (Figure~\ref{fig:shape}G). The corresponding Raman spectra are presented in Figure~\ref{fig:shape}E. The spectral shift between the lowest (AR = 1) and the highest (AR = 5) aspect ratios reaches up to $\sim$ 6 cm$^{-1}$, accompanied by a reduction in the SPhP-to-TO intensity ratio with increasing AR (Figure~\ref{fig:shape}H).

The SPhP/TO ratios, shown in Figure~\ref{fig:shape}D,H for dimers and ellipsoids, respectively, decrease with reduced gap size and increased aspect ratio. This further supports the localization condition; as dimer gaps reduce, the structure behaves more like a single, larger object, while increasing AR similarly increases the nanopillar size. In both cases, the system approaches the excitation wavelength, thereby suppressing efficient Raman excitation. This is particularly pronounced for AR = 4 and 5 ellipsoids, where the SPhP/TO ratio drops to 0.05 -- about 4-fold lower than that of AR = 1.

More complex structures, such as plasmonic-inspired heptamers designed to mimic Fano resonances, are presented in Figure S6. Importantly, although Fourier transform infrared measurements may reveal complex phenomena like Fano interference, the Raman response is largely featureless (see Figure S6). This difference is likely due to the spatial localization of SPhPs relative to the excitation wavelength, with the heptamer structures being relatively large compared to Raman $\lambda_{\rm exc}$ in the visible range, but small relative to that in the infrared.

\begin{figure}%[ht!]
    \centering
    \includegraphics[width=0.75\linewidth]{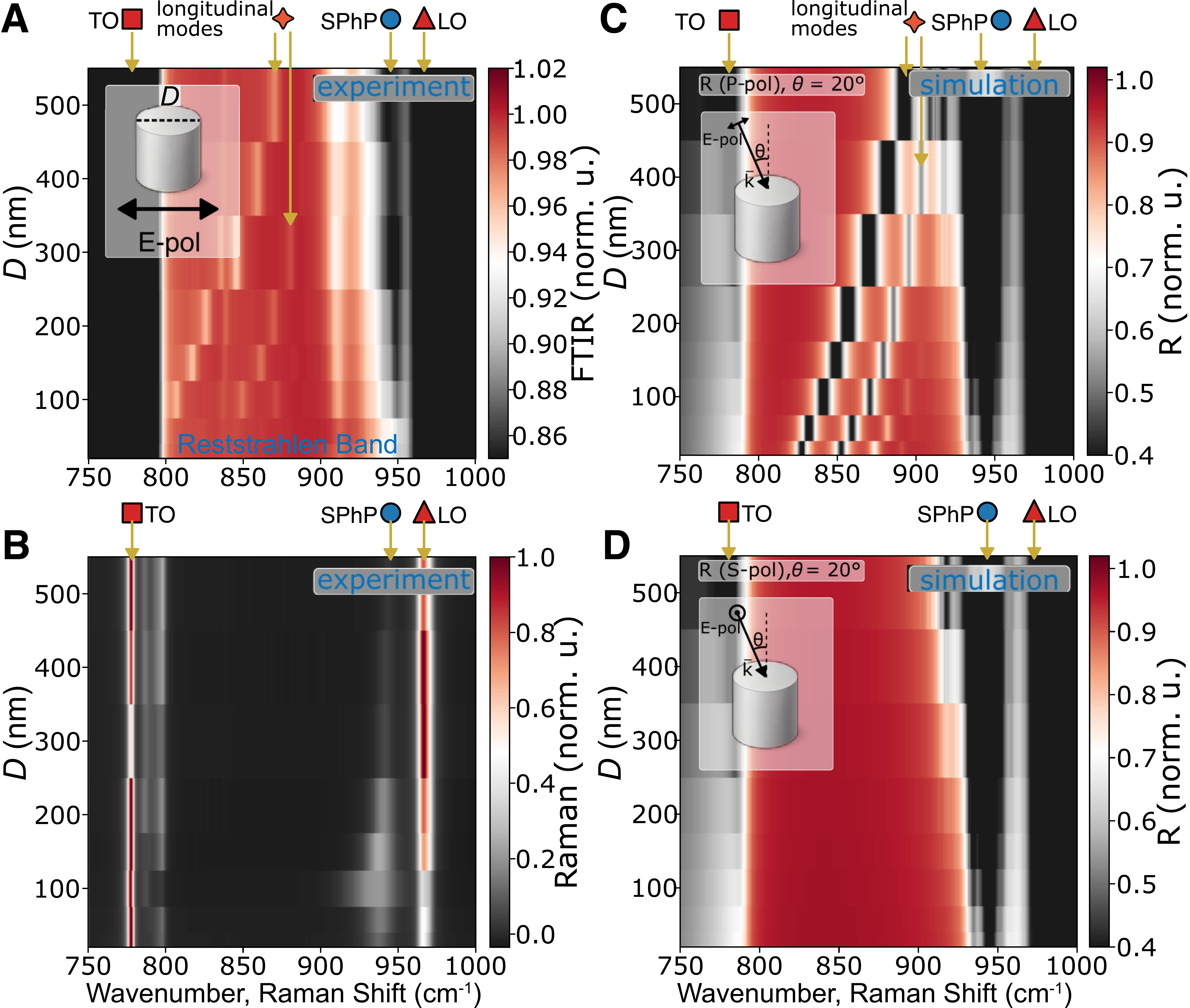}
    \captionsetup{font=scriptsize}
    \caption{\textbf{Comparison of infrared and Raman spectra for SiC nanopillars of various diameters.} \textbf{A}, Infrared reflection spectra of a pillar array of various diameters from 30 to 500 nm. The center-to-center pitch size is 2 $\mu$m. The inset is the schematic of the experiment. The arrows and markers indicate the position of the respective modes -- TO, LO (indicates the border of the Reststrahlen band), SPhP, and lattice resonances from the periodic nature of the sample and mid-IR excitation. \textbf{B}, Experimental Raman scattering of the single nanopillar array from \textbf{A}. The markers and arrows indicate the same set of modes, but the longitudinal resonances are absent in the Raman scattering. \textbf{C}, Numerically simulated reflection map of SiC pillar array on the SiC substrate. The incident light is a $p$-polarized plane wave excitation with a $\theta$=20$\degree$ angle with respect to the normal. \textbf{D}, Numerically simulated reflection map of SiC pillar array on the SiC substrate. The incident light is a $s$-polarized plane wave excitation with a $\theta$=20$\degree$ angle with respect to the normal. The parameters of simulated SiC pillars closely approximate the experimental ones.  The markers and arrows indicate the same set of modes, but the lattice resonances are not present in the Raman scattering.}
    \label{fig:FTIR}
\end{figure}

\paragraph*{Comparison of Raman and infrared spectra of phonon polaritons.} Correlating the infrared and Raman spectroscopy data of nanopillar structures serves two main goals: (\textit{i}) it enables the identification of modes that are active in the infrared versus Raman spectra, allowing for a comparison of their selection rules, and (\textit{ii}) it allows us to examine whether modes that appear in both Raman and infrared spectroscopies exhibit similar scaling behavior with respect to structural parameters such as nanoparticle size, dimer gap, and aspect ratio. Structural scaling is well-known in plasmonics within the framework of optical nanoantennas~\cite{novotny2007effective} and therefore can serve as a valuable reference.

By comparing Raman and infrared spectra of SiC nanopillars, we identify a striking anti-correlation: as nanopillar diameter ($D$), inverse dimer gap ($\delta^{-1}$), or aspect ratio (AR) increase, the SPhP peak blueshifts in Raman spectra, contrary to plasmonic expectations~\cite{novotny2007effective,gunnarsson2005confined}, but redshifts in infrared (Figure~\ref{fig:FTIR} and Figure S7). This contrast persists across single pillars, dimers, and elliptical shapes, suggesting a fundamental difference in the way the two spectroscopic techniques probe phonon polaritons (Figure S7). Moreover, such anti-correlated behavior is consistent with that observed in an alternative phonon polariton material platform, gallium nitride (GaN), the results for which are presented in Figure S11.

Raman selectively excites transverse, while infrared -- both transverse and longitudinal pillar modes, indicating that Raman excitation requires not only the presence of Raman- and infrared-allowed SPhPs (Figure~\ref{fig:FTIR}), but also their localization on a length scale smaller than the excitation wavelength, $D < \lambda_{\rm exc}$. The absence of longitudinal modes in Raman and the opposite frequency trends indicate differences in selection rules and excitation mechanisms. A simple longitudinal-transverse mode explanation for the anisotropically shaped SiC nanopillars meets quantitative challenges, as both Raman and infrared peaks appear approximately in the same spectral range, despite significantly different sensitivities of transverse and longitudinal resonances to the structure of the pillars predicted theoretically (SI theory, Figure S9). The observed anti-correlation thus likely arises from the interplay between particle size and anisotropy, material birefringence~\cite{wang2017phonon}, and excitation conditions. Data for elliptical SiC particles is presented in Figure S8.

\paragraph*{SiC nanostructures with subwavelength dimensions in all three directions.} So far we focused on SiC pillars that satisfy the $L > \lambda_{\rm exc} > D$ condition. This may explain the absence of longitudinal SPhPs modes in the Raman spectra of such structures, since their confinement is mismatched with the excitation wavelength in the vertical direction. Such mismatch could be illustrated by viewing the Raman polarization as a product between the SPhP eigenmode $Q_\mathrm{SPhP}$ and the excitation electric field $E_0$, $p_{\rm Ram}(\lambda_{\rm Ram}) \propto \left(\frac{\partial \alpha}{\partial Q_{\rm SPhP}}\right)_0 \cdot Q_{\rm SPhP}(\lambda_{\rm SPhP}) \cdot E_0(\lambda_{\rm exc})$, and noting that these quantities are localized on the nanopillar and excitation wavelength scale, respectively. When the particle dimensions are larger than the excitation wavelength, polaritonic Raman dipoles coherently and constructively interfere only in the near-forward scattering direction, which reproduces the bulk polariton behavior~\cite{henry1965raman}. In contrast, when the particle dimensions are smaller than the excitation wavelength, these restrictions do not apply, and Raman dipoles radiate in all directions, making them detectable in the back-scattering configuration used in our experiments. Therefore, to be observable, these SPhP modes must be confined below the excitation wavelength along all three dimensions, $\lambda_{\rm exc} > L, D$. Furthermore, it is important to note that Raman scattering of SPhPs has previously been reported only at frequencies close to LO phonons~\cite{hayashi1982raman,hayashi1985raman,caldwell2013low}. It is important to clarify whether this limitation is fundamental.

%To shed light on this problem, we engineer anisotropic and subwavelength in all three dimensions SiC nanostructures.
%Indeed, a Raman response deep in the Reststrahlen band may open doors for regimes of interactions.
%Such a comparison allows us to conclude about the generally correct understanding of SPhPs excitation in the Raman process (alternatively, a new theory has to be developed).

\begin{figure}%[ht!]
    \centering
    \includegraphics[width=0.95\linewidth]{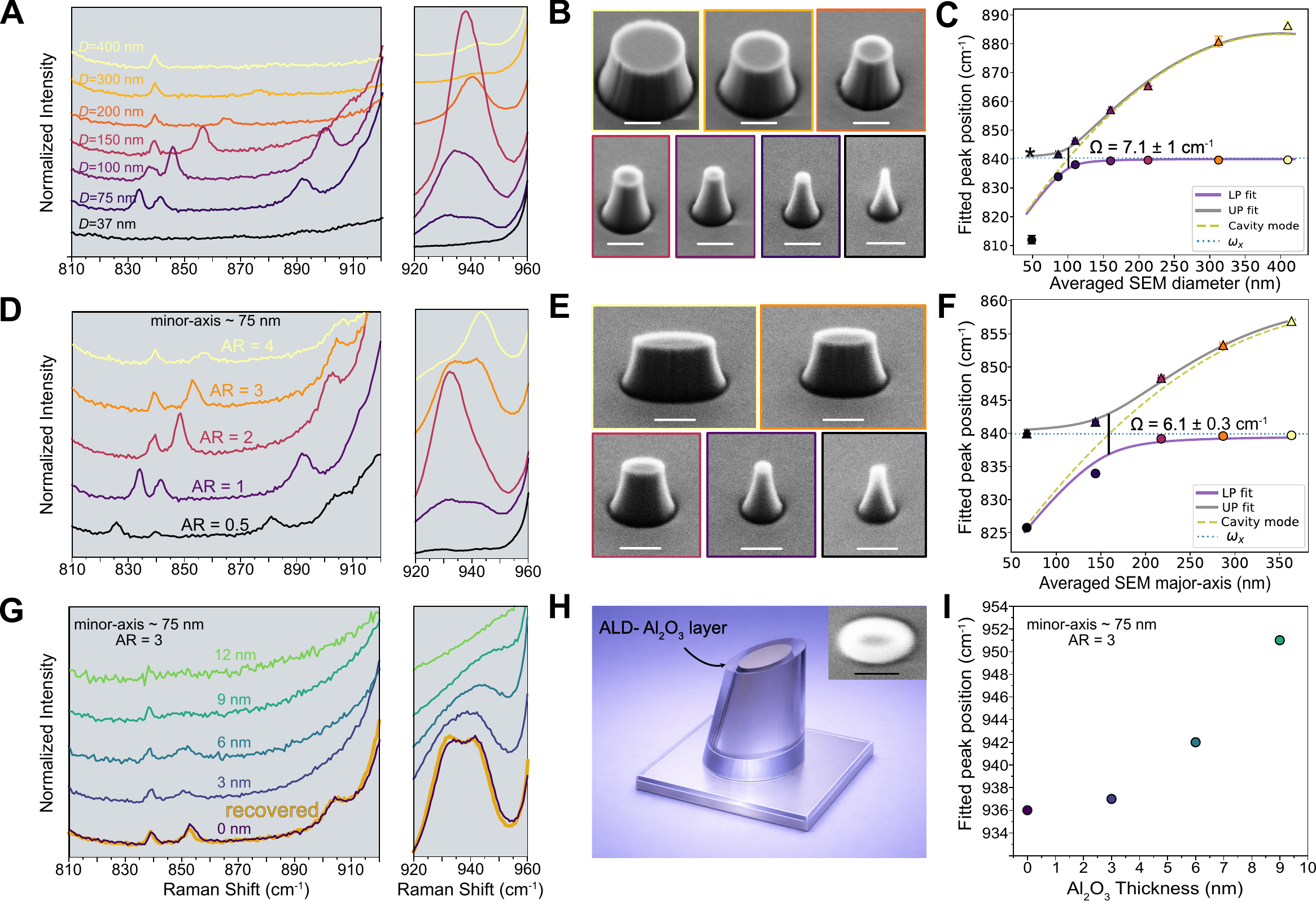}
    \captionsetup{font=scriptsize}
    \caption{\textbf{Raman scattering by surface phonon polaritons in subwavelength 4H-SiC nanostructures.} \textbf{A}, Raman spectra for different diameters of nanopillars with $L=270$~nm. Diameters given in the legends are the design diameters. \textbf{B}, Tilted-view SEM images of SiC nanopillars. Scale bars are 200~nm. \textbf{C}, Fitted peak position of Raman scattering spectra for corresponding pillars. Diameter values on the $x$-axis represent the average of the top and bottom diameters of the nanopillars, extracted from the corresponding SEM images. Error bars indicate the standard deviation calculated from six to nine pillars for each diameter. [$\star$] symbol marks an unidentifiable peak position, caused by the limited fitting accuracy in this case, due to the low signal-to-noise ratio in the Raman spectrum. For some diameters, error bars are smaller than symbols. \textbf{D}, Typical Raman spectra for elliptical nanopillars with minor axis $\approx$ 75~nm and height of 270~nm. Aspect ratio is shown as AR = 1 -- 4. \textbf{E}, Tilted-view SEM images for elliptical nanopillars with minor axis $\approx$ 75~nm. Scale bars are 200~nm. \textbf{F}, Fitted peak position of Raman scattering spectra for elliptical nanopillars. The major axes on the $x$-axis represent the average of the top and bottom major axes of the nanopillars, extracted from the corresponding SEM images. Error bars indicate the standard deviation calculated from six to nine pillars for each elliptical nanopillar. For some pillars, error bars are smaller than symbols. \textbf{G}, Effect of different thicknesses of  Al$_2$O$_3$ (0-12~nm) on Raman spectra of elliptical nanopillars with aspect ratio 3 (AR = 3)  and height 270~nm). Orange spectrum shows the recovery of the SPhP peaks after cleaning the Al$_2$O$_3$ layer.   \textbf{H}, Graphical image of elliptical nanopillar. Inset is the top view SEM image of the particle minor axis $\approx$ 75~nm and AR = 3 with 3~nm Al$_2$O$_3$ deposited by ALD. Scale bar is 200~nm. \textbf{I}, Fitted peak position of SPhP with varying Al$_2$O$_3$ thicknesses for the presented spectra in G. The data corresponding to a 12~nm Al$_2$O$_3$ thickness is excluded because the SPhP Raman peak is poorly defined and shows a pronounced blue shift with increasing Al$_2$O$_3$ thickness. }
    \label{fig:subwavelength}
\end{figure}

% modes deep in Reststrahlen
To shed light on the matter, we produced SiC nanostructures that are subwavelength in all three dimensions. The results, presented in Figure~\ref{fig:subwavelength}, indicate the appearance of a rich Raman mode structure. In particular, additional SPhPs modes appear in the 800 -- 900 cm$^{-1}$ range, deep in the Reststrahlen band, \textit{i.e.} at frequencies significantly below the standard SPhPs in larger structures. All structures in Figure~\ref{fig:subwavelength} have the same height of $L=270$~nm. Cylindrical particles are summarized in A-C, while ellipsoidal structures with AR = 1 -- 4 are shown in D-F for a 75-nm minor axis nanoparticle (additional data for a 37-nm minor axis nanoparticle are shown in Figure S16). All particles clearly demonstrate dispersion as the particle size and aspect ratio are varied, supporting the polaritonic nature of these modes.

%All data clearly demonstrates additional SPhP modes in the 800 -- 900 cm$^{-1}$ spectral range, in the range of lower frequencies below the standard SPhPs in larger structures, and thus lying deep in the Reststrahlen band.  

% dependence on particle volume

 %(\red{or it is a volume squared? in fact important to figure out. For incoherent molecular ensembles $I_{Raman} \propto N \propto V$. In a coherent system, one would rather expect $V^2$ scaling. Which one do we have? Need to analyze the data and plot it as a separate panel})

It is important to note that the subwavelength SPhP mode localization comes with a trade-off. Reducing the particle dimensions enhances the efficiency of Raman excitation. At the same time, the overall scattering signal scales with the volume of the nanoparticles. The design must balance the requirement of subwavelength localization with that for a sufficiently large particle volume to preserve a measurable Raman response (Figure~\ref{fig:subwavelength}). Therefore, an optimal nanoparticle size is anticipated, consistent with the observations presented in Figure~\ref{fig:environment}A–D and Figure~\ref{fig:subwavelength}. Importantly, the optimal size range is relatively narrow: for instance, for a given $L=270$~nm, the optimal diameter of cylindrical pillars lies between 40 and 400 nm, with the minimum size set by the threshold for sufficient Raman signal, while the maximum determined by subwavelength confinement (Figure~\ref{fig:subwavelength}A). The precision required for both parameters explains why these results are challenging to achieve and why the previous attempts (especially earlier works conducted on disordered ensemble systems) might have missed the effect (specifically, the second Raman SPhP condition (\textbf{2}) is violated by our findings). Furthermore, it is important to clarify that Raman selection rules result in a much narrower range of parameters ($L, D, \lambda_{\rm exc}$) under which a Raman signal can be observed, which is in stark contrast with infrared spectroscopy, where the subwavelength localization condition is relaxed due to the longer excitation wavelength of infrared photons.

%Importantly, the optimal size range turns out to be relatively narrow, i.e. for the optimal $D$ (for a given $L=270$~nm) seems to be in between 40 and 400 nm, so the range of the effect is not very broad! This may explain why it is so hard to obtain these results and why it has not been observed previously. Conditions on both $L$ and $D$ are rather precise. 

%\red{Panels with anticrossing, consider adding vertical lines for bare phonons, and curved lines for optical modes. Also, this, in principle, should include coupled oscillator model analysis. Align panels. In panel C, a pair of data points for D=37 nm is missing.}

% self-hybridized anticrossing
%\red{(ref)}

Interestingly, we observe a localized phonon mode at approximately 840 cm$^{-1}$. This mode arises from pure phonon vibrations, and is also known as a weak phonon mode, which appears at 839 cm$^{-1}$ in 4H-SiC~\cite{bluet1999weak}. However, SPhPs in subwavelength nanoparticles appear in close spectral proximity to this weak phonon mode, and when tuned sufficiently near resonance, they hybridize with it (Figure~\ref{fig:subwavelength}A,C,D,F). This represents an intriguing regime of self-hybridization, as such self-hybridized polaritons and their associated anticrossing in Raman scattering have not, to our knowledge, been reported previously. The observed anticrossing behavior can be quantitatively described using a coupled oscillator (see SI for quantification of the anticrossing). The corresponding results are presented in Figure~\ref{fig:subwavelength}C,F, for cylindrical and ellipsoidal particles, respectively. Furthermore, this anticrossing can, in principle, be tuned by modifying the electromagnetic environment, as demonstrated by the deposition of a thin carbon layer during SEM experiments (Figure S17).

% sensing with ALD
Finally, to verify the sensing capability of individual subwavelength SiC structures, we sequentially coated them with conformal alumina (Al$_2$O$_3$) layers deposited by atomic layer deposition (ALD) until the response was saturated. The thickness of each alumina layer in the sequence was accurately controlled and fixed at 3~nm. Even the deposition of only a few nanometers of alumina induces clear shifts in the Raman modes. This observation is particularly intriguing, since the resonant wavelength of SPhPs is in the mid-infrared, where the free-space wavelength reaches several micrometers -- orders of magnitude greater than the thickness of the deposited layer. This observation is thus promising for applications in refractive-index sensing and thin-film deposition. Notably, the Raman modes do not universally redshift, as is typically expected in conventional plasmonic refractive-index sensing and also as shown in Figure~\ref{fig:environment}E-H for taller SiC nanopillars immersed in various liquids. Instead, in Figure~\ref{fig:subwavelength}G-I, the main SPhP mode (close to LO) shows a blue shift. This behavior follows the refractive index dispersion of alumina in the mid-infrared and demonstrates that, although the sensing signal is displayed in the visible by Raman spectroscopy, it is intrinsically sensitive to the optical response of the material in the mid-infrared.

\section*{Conclusions}
In conclusion, we demonstrate the detection of Raman scattering by surface phonon polaritons in individual SiC and GaN nanopillars and their arrays, exhibiting a clear dispersion with respect to the nanopillar size, shape, arrangement, and surrounding electromagnetic environment. This enables refractive index sensing in the mid-infrared range using visible-wavelength optical equipment, potentially benefiting from a richer span of dielectric functions accessible in the fingerprint region. We demonstrate a crucial difference between Raman and infrared -- phonon polaritons must be localized on scales smaller than the excitation wavelength in all three dimensions, and since that is much shorter in the visible range, only relatively small nanostructures can be efficiently excited in Raman scattering. We stress that SiC offers excellent chemical and mechanical stability, wafer-scale production, as well as a high melting point, compatible with harsh experimental conditions, and enabling repeated use. Furthermore, the ultrasharp nanopillars studied here (with $D \sim 10$ nm) may be promising for use as nanoscale scanning probe microscopy tips, combining durability with nanoscale resolution and environment-specific Raman readout, which additionally could integrate vacancy centers in SiC in a scalable manner~\cite{koehl2011room,lukin20204h} or include thermal emission management~\cite{schuller2009optical}. Furthermore, SiC and GaN are both wide bandgap semiconductors used in multiple applications, including high-power electronics~\cite{buffolo2024review}, light-emitting diodes, and edge-emitting lasers, where SPhP Raman spectroscopy of nanostructured features may enable advanced device characterization, such as contamination monitoring or thin-film deposition. With the advancement in monolithic SiC metasurfaces~\cite{schaeper2022monolithic,chen20254h}, the present technology can be made even more efficient in terms of resonant excitation of the Raman scattering process (exemplified in Figure S20 through observation of pronounced reflection colors and in Figure S21 through fabrication of various plasmonics-inspired SiC nanostructures). Finally, in this work, we strengthen the analogy between SPhPs and VSC by highlighting the structure-dependent behavior observed in both platforms, and hypothesize that Raman spectra under VSC should be similarly responsive to environmental changes as well as require subwavelength polariton localization~\cite{dherbecourt2025spontaneous}.

%Moreover, dense SiC metasurfaces may produce an additional optical response in the visible range, providing a secondary sensing channel (as exemplified in SM, \red{color pictures of SiC dense arrays should be shown}). This opens the door to a dual-mode sensing platform that combines Raman-based detection with metasurface-enhanced signals. 

%Compared to traditional plasmonic sensing platforms, the proposed SiC-based scheme offers an innovative Raman-based sensing approach, enhanced robustness, and potential reusability. 

\section*{Materials and methods summary}

Nanofabrication was carried out at the Myfab Nanofabrication Laboratory, MC2 Chalmers. Raman and FTIR spectra, as well as SEM and TEM measurements, were performed at the Chalmers Material Analysis Laboratory (CMAL). The samples were fabricated using electron-beam lithography, followed by dry etching and cleaning. A detailed description of the Methods used is provided in the SI.

\bibliography{main}
\bibliographystyle{science}

\section*{Acknowledgments}
We thank Dr. A. Canales for help at an early stage of the project. We additionally thank Prof. R. Yakimova for providing the 6H-SiC substrate and Prof. A. Kalaboukhov for providing carbon deposition for FIB  of SiC pillars. \textbf{Funding:} T.O.S. acknowledges funding from the Swedish Research Council (VR project, grant No. 2022-03347), Chalmers Area of Advance Nano, 2D-TECH VINNOVA competence center (Ref. 2024-03852), Olle Engkvist foundation (grant No. 211-0063), and the Knut and Alice Wallenberg Foundation (KAW, grant No. 2019.0140). The Swedish Research Council and Swedish Foundation for Strategic Research are acknowledged for access to ARTEMI, the Swedish National Infrastructure in Advanced Electron Microscopy (2021-00171 and RIF21-0026). E.L. acknowledges funding from C3NiT-Janzen VINNOVA Competence Center (Grant No. 2022-03139) as well as from NanoLund. This work was performed in part at Myfab Chalmers and the Chalmers Material Analysis Laboratory (CMAL). \textbf{Authors contributions:} T.O.S. conceived the project idea. G.Z. and B.K. performed optical measurements. B.K. fabricated the SiC samples with the help of N.S. and S.L.A. A.N.S. and E.L. designed and fabricated GaN samples. Transmission electron microscopy experiments were performed by L.Z., A.B.Y., and A.R. G.Z. and T.J.A. calculated infrared spectra of SiC pillars and their arrays. O.K. performed analytical calculations of spheroidal SiC pillars. G.Z., B.K., and T.O.S. wrote the manuscript with contributions from all co-authors. E.O., S.L.A, and T.O.S. supervised the study.
\textbf{Competing interests:} The authors declare that they have no competing interests.
\textbf{Data and materials availability:} Experimental procedures and characterization data are provided in the supplementary materials. Correspondence and requests for materials should be addressed to T.O.S. (timurs@chalmers.se).
%\textbf{License information:}

\section*{Supplementary materials}
This material contains a detailed Methods section, as well as additional figures and notes, supporting the findings of the main text.

\end{document}